# Observation of one electron charge in an enhancement-mode InAs single electron transistor at 4.2K

*G. M. Jones, B. H. Hu, C. H. Yang*

Department of Electrical and Computer Engineering, University of Maryland

College Park, MD 20904

*M. J. Yang*

Naval Research Laboratory, Washington DC 20375

*Y. B. Lyanda-Geller*

Department of Physics, Purdue University, West Lafayette, IN 47907

*Abstract*

We propose and demonstrate experimentally a novel design of single-electron quantum dots. The structure consists of a narrow band gap quantum well that can undergo a transition from the hole accumulation regime to the electron inversion regime in a single-top-gate transistor configuration. We have observed large size quantization and Coulomb charging energies over 10meV. This quantum dot design can be especially important for scalable quantum computing.

Few and one-electron quantum dots (QDs) are artificial atoms that are of fundamental importance to modern science and technology.[1] The possibility to connect the dots to electron reservoirs provides an excellent tool for investigating the atomic-like properties of the dots and using them in optoelectronics. Creating QDs with just one electron present is of particular interest. Such QDs are prospective candidates for devices generating single photons.[2,3] They are also of great interest for rapidly developing field of solid state quantum computing. One electron in a QD represents a single spin qubit that has been proposed[4] as the building block for implementing a quantum computer. Being able to create, confine, manipulate, and probe a single spin is vital for quantum information processing.[5]

Identifying a single electron charge was first demonstrated[6] by using vertical tunneling single electron transistors (SETs), in which a double barrier tunneling heterostructure is mesa-etched into a cylindrical pillar. Electrons in this setting are confined in the $In_xGa_{1-x}As$ QD, and the epitaxial $Al_xGa_{1-x}As$ tunnel barriers define their probability of tunneling to the source/drain regions. A Schottky gate surrounding the sidewall of the pillar is biased to deplete the number of electrons inside the vertical quantum dot electrostatically. Early experimental progress and understanding of the single electron addition energy spectrum[7] was achieved in this configuration. There were also successful observations of a single electron charging on lateral SETs.[8,9] Based on the versatile high mobility two-dimensional (2D) electron gas in AlGaAs/GaAs single heterojunctions, patterned top Schottky gates are biased to deplete electrons beneath them and form an SET with one electron in the QD. Lateral SETs provide the experimental system that led to many significant observations in the few- electron regime.[8,9]

Vertical and lateral QDs can be used not only for creating single qubits, but also for coupling into quantum gate.[10,11] Exchange coupling between two neighboring spins has been investigated in both types of SET. It has been mathematically proven that combining quantum gate operations, such as mutual spin flip of two qubits (i.e., swap) and single qubit spin rotations is sufficient for performing any quantum computation. However, comparing the prospects for scalable solid state quantum computing using vertical and lateral SETs, it is clear that the outlook for lateral SETs is more encouraging, particularly in the light of success in industrial MOSFETs (metal oxide semiconductor field effect transistors) scaling. Unfortunately, conventional depletion-mode lateral SETs have a significant drawback: their tunnel barriers tend to rise when the plunger gates are emptying the QDs, and that could totally block the current flow when QDs are still in many-electron regime. Inserting additional quantum point contact in the proximity of SET can circumvent this problem by bringing an ability to identify a single electron, but these additional point contacts make the scaling architecture complicated.

In this paper, we propose and demonstrate an *enhancement-mode* lateral single electron transistor. In contrast to the *depletion-mode* SETs that reach one-electron regime by expelling electrons from multi-electron QDs, the enhancement-mode SET contains no electrons initially. Our SET structure uses a single top gate to create two symmetric tunnel barriers and make electrons tunnel into this empty quantum dot one at a time, as evidenced by the current-voltage characteristics of the SET. This enhancement-mode setting possesses the advantages of both vertical and lateral depletion-mode SETs. First, the tunneling barriers that define a quantum dot are, to the leading order, independent of the top gate voltage in the few-electron regime. Second, lateral configuration makes the control of the coupling of two dots straightforward. In addition, the simplicity of a single top gate configuration makes possible the realization of a 2D QD that would be necessary to allow transportation of qubits.[12]

The samples used in this work to demonstrate this novel SETs concept are InAs/GaSb composite quantum wells (CQWs)[13] grown by molecular beam epitaxy on (001) undoped GaAs substrates. The InAs/GaSb heterojunction has a staggered energy band alignment, as shown in Figure 1 (a). The CQW is sandwiched by $Al_xGa_{1-x}Sb$ barriers, and the bottom $Al_xGa_{1-x}Sb$ is modulation p-doped. That is, in the as-grown samples, the GaSb QW (16nm) is populated with 2D holes, and the InAs QW (4nm) is free of electrons. Effectively, the CQW becomes a narrow-bandgap semiconductor, whose bandgap is tunable by the thicknesses of the two QWs and is ~100meV in our case. The sample is fabricated into the transistor structure by either the standard photolithography or electron-beam lithography.

Figure 1 (b) shows the schematic of the transistor structure. The top gate voltage ($V_{gate}$) controls the population and the type of carriers in the CQW. At $V_{gate}$=0, the transistor conducts because of the 2D holes in the GaSb layer. As the top gate is biased to be more positive, the 2D holes in the GaSb QW are gradually depleted. Further increase of the top gate bias results in the formation of electrons in the InAs layer. Such gate-controlled transition from accumulation of 2D holes to depletion and finally to inversion of electrons is verified by large sized gated Hall bars (with 10 μm channel width and 100 μm channel length). The dc current-voltage characteristics measured at 4.2K in the common-source configuration is plotted in Fig. 2, where the potential profiles along the source-drain direction in three operating regimes are also shown. The conductance is found to be large when there are either 2D electrons or 2D holes in the CQW under the gate area. Both the gate leakage current and the drain current in the depletion region are less than the measurement resolution (0.1pA).

In order to understand the effect of gating better and to provide the parameters for the design of the SET devices, we have computed numerically the potential profile of InAs QDs.

The model structure, as depicted in the left inset of Fig. 3, is a standard MOS capacitor, but the gate here is a small metallic cylinder with a diameter (*D*) less than 100 nanometers. Details of the numerical simulation are discussed elsewhere. The right inset of Fig. 3 shows an example of the simulation assuming the effective band gap of the CQW being 100meV and a 2D hole density of $1\times10^{12}$ cm$^{-2}$ in the GaSb layer. We find that for a smaller gate diameter, although it takes a higher gate voltage to induce the first electron, the confinement potential is steeper and the size-quantization energies are larger. For example, when a single electron resides in the QD, the quantization energy is 13meV, 15meV and 17meV for *D* = 100nm, 75nm and 50nm, respectively. One key operating principle, substantiated through the simulation, is that the band-to-band tunneling barrier is not sensitive to the top gate bias for the first few electrons, as illustrated in Fig. 3. The height of the tunneling barrier and its width are primarily dependent on the band gap of CQW and 2D hole density, respectively. In other words, the tunneling transmission coefficient in enhancement-mode QDs is determined by the potential profile of the heterostructure, similarly to the depletion-mode vertical QDs.

For the fabrication of SETs, we have used electron-beam lithography and wet etching, to fabricate transistors with a gate length down to a few tens of nanometers. Transistors with a channel width of approximately 700nm show a complete depletion of 2D holes, indicating a surface depletion of 350nm at the sidewall. The small transistors display the expected Coulomb blockade. Fig. 4 shows the data taken at 4.2K, where the dc drain current is measured at a fixed drain voltage of 1mV, and the gate voltage is swept from zero to 9 volts. At $V_{gate}$=0, the system is in the accumulation regime, and the transistor shows ohmic conductance. As the gate voltage increases, the drain current drops because the transistor goes from accumulation to depletion regime, where the drain current is zero. When the gate voltage bias is higher, 6.2V, the lowest single electron quantum state beneath the gated region is aligned with the Fermi level, resulting in a current peak. As the gate voltage becomes more positive, the current peaks due to the second and the third electron occupation are observed near $V_{gate}$ = 7.0 V and 8.4 V, respectively.

To investigate the operation of this enhancement mode InAs SET further, we have carried out the stability measurement: the drain current is measured as a function of a stepping gate voltage and a sweeping drain bias. Fig. 5 shows the contour plot of the drain current using linear scale. In manifestation of single electron transport in the Coulomb blockade regime, the contour plot shows a series of diamond-shaped blocks. The dashed lines highlight the boundaries of the blocks and are labeled by N = 0, 1, 2, 3, and 4, referring to the number of electrons in the QD. According to the "orthodox" theory[14] of the Coulomb blockade, the height ($e/C_\Sigma$), the width ($e/C_{gate}$), and the slopes defining the diamond ($C_{gate}/C_{drain}$ and $-C_{gate}/C_{source}$) uniquely determine the SET charging energy, where $C_{gate}$, $C_{drain}$, and $C_{source}$ are the capacitance between the quantum dot and the gate, the drain, and the source, respectively, and $C_\Sigma = C_{gate} + C_{drain} + C_{source}$. Following this standard analysis procedure, we have calculated for the 1st, the 2nd, and the fourth diamonds respective capacitances and the "addition energies", as shown in Table 1. We obtained an addition energy of 15meV for the first block, 35 meV for the 2nd, indicating that the level spacing due to size quantization effect is 20meV. Comparing to our potential simulation, the obtained quantization energy implies a QD less than 50nm in diameter. If we model the QD as a disc with a radius *r*, the obtained capacitance of our QD suggests an effective diameter of about 20nm. Here, we use $C_\Sigma = 8\varepsilon r$ and $\varepsilon$=12.3 for the dielectric constant of InAs. We notice that the 3rd diamond contains several spikes. We attribute them to single electron traps in the vicinity of SET that happen to be activated under specific operating conditions. The size of the 3rd diamond should therefore be smaller than the apparent result shown in Fig. 5.

The observations of relatively large Coulomb and size-quantization energies are significant, and we attribute these features to the novel design of the InAs enhancement mode SET. The criteria for observing single electron tunneling characteristics are that the Coulomb energy ($e^2/C_\Sigma$) be larger than the thermal energy, and that the tunneling conductance be smaller than the conductance quantum $e^2/h$. Consequently, SETs in the many-electron regime have been demonstrated in a variety of systems, including $In_xGa_{1-x}As$, GaAs, Si, carbon nanotubes, and metal-based structures. However, the experimental requirement is more stringent for observing the size quantization effect, due to two additional criteria: One is that the Fermi wavelength should be comparable to the island's size, and the other is that the number of electrons on the island should correspond to the few-electron regime. Depletion-mode SETs were the only systems reported thus far that show size-quantization energies. Typically the size quantization energy of depletion-mode SETs is a few meV, while our enhancement mode InAs QD exhibits size quantization energies an order of magnitude larger. Larger quantization energy allows for SET operation at a higher temperature, as evident from 4K operating temperature of our SET versus 10-50mK in depletion-mode SETs.[6][9] Large quantization energy is important for quantum computing: The spin decoherence caused by admixture of electron states due to spin-orbit interaction is reduced as a result of large orbital level energy spacing, making the system closer to an ideal two-level system. The observed large quantization energy is due to small electron effective mass in InAs and a steep confinement potential resulting from the band-to-band tunneling design. We note that the steep confinement potential makes it possible to place QDs in close proximity needed for stronger exchange coupling between dots and for constructing an array of QDs capable of transporting qubits via the paths of empty QDs.

The straightforward identification of a single electron occupying the QD is a significant factor that has important implication for the implementation of qubits. Similarly to vertical InGaAs QDs, in our enhancement SETs this verification is achieved by QD conductance measurement. As our data indicates, the voltage difference between the cutoff of the accumulation and the onset of the inversion is about 3V. This onset is repeatedly measured in transistors of different sizes on the same wafer. It is determined by the structure and by the fixed p-doping concentration. The first occurrence of a current peak in the SET characteristics is around 3V above the cutoff of the accumulation as well, suggesting that it arises from the electron ground state of the QD. Furthermore, the subsequent conductance peaks are similar in amplitude, as it should be for band-to-band tunneling process. Moreover, the difference in gate voltage between the neighboring peaks is on the order of 1V. It is extremely unlikely that the first peak near $V_{gate} = 6V$ comes from resonant tunneling through, say, the $10^{th}$ electron state: If this is the case then the current peak N=9 should be observed near $V_{gate} \sim 5V$ as well. Because the tunneling peaks are similar in magnitude, any resonance at $V_{gate} < 6V$ would have been observable. We therefore conclude that the observed peak near 6V corresponds to the lowest single electron state in the InAs QD.

In conclusion, we have proposed an enhancement-mode SET and demonstrated it using InAs/GaSb composite QW that gives a QW with a narrow band gap. We achieved the one-electron regime in InAs quantum dot with a size quantization energy of 20meV. Our approach is advantageous for applications in quantum information technology, including lateral configuration, a-single-top-gate design, steep potential confinement, and straightforward identification of a single electron.

Acknowledgements: This work is supported by NSA/ARDA, LPS/NSA and ONR.

**Figure captions**

Fig. 1. (a) Schematic of band diagram of the composite quantum well. (b) Schematic of the enhancement-mode single electron transistor, where one electron is induced in the InAs layer by a top metal gate.

Fig. 2. The dc current-voltage characteristic of a gated Hall Bar transistor. The insets plot the schematic potential profiles along the current direction for three operating regimes, where $E_c$ ($E_v$) is the first electron (hole) subband in the InAs (GaSb) layer.

Fig. 3. Calculated potential profiles when the first (dashed curve, where the center of the dot is shifted for comparison of the tunneling barrier) and the second quantization level (solid) are aligned with the Fermi level. Here the Coulomb charging energy is excluded and the shaded areas illustrate the tunneling barriers. The insets show the model capacitor structure and a simulated 2D potential plot.

Fig. 4. (a) The dc current-voltage characteristic of a gated single electron transistor, where the drain current is plotted against the sweeping gate voltage. The inset shows a scanning electron micrograph of a transistor, where the scale bar is 1 μm in length. (b) Same as that in (a), but with a different drain current scale, where the peaks in current results from single electron tunneling.

Fig. 5. Diamond chart of the enhancement-mode SET.

Table 1. The capacitances and the addition energies obtained from the diamonds in Fig 5.

| N | 1 | 2 | 4 |
|---|---|---|---|
| $C_{gate}$ | 0.69 aF | 0.29 aF | 0.37 aF |
| $C_{drain}$ | 4.7 aF | 1.7 aF | 5.1 aF |
| $C_{source}$ | 5.4 aF | 2.6 aF | 4.4 aF |
| $E$ | 15 meV | 35 meV | 16 meV |

Fig. 1, Jones et al.

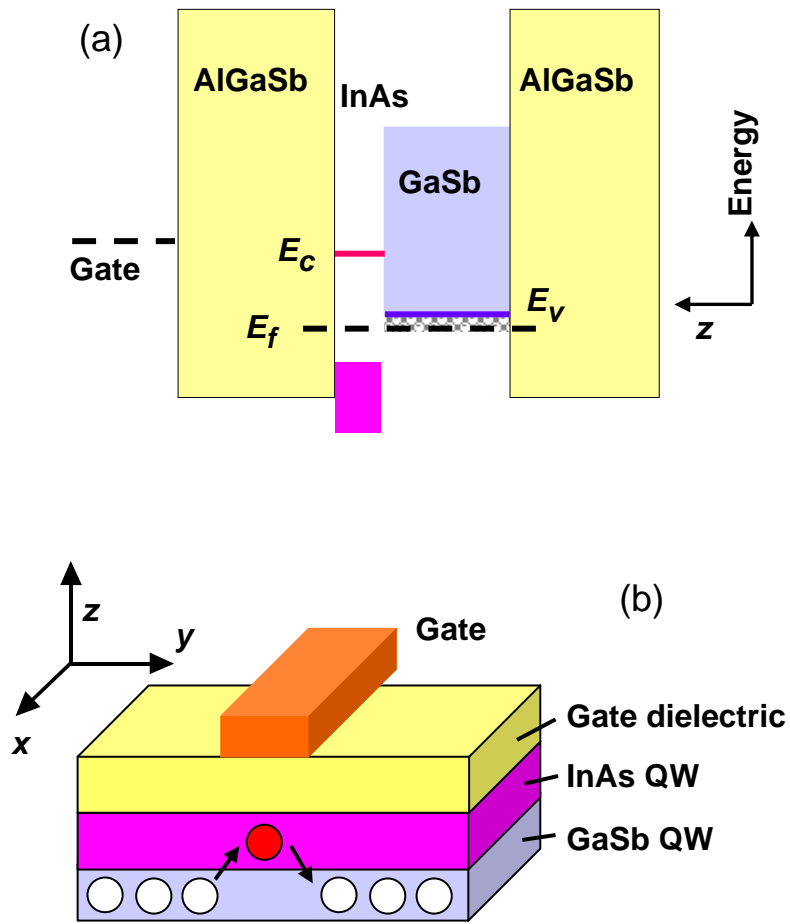

Fig. 2, Jones et al.

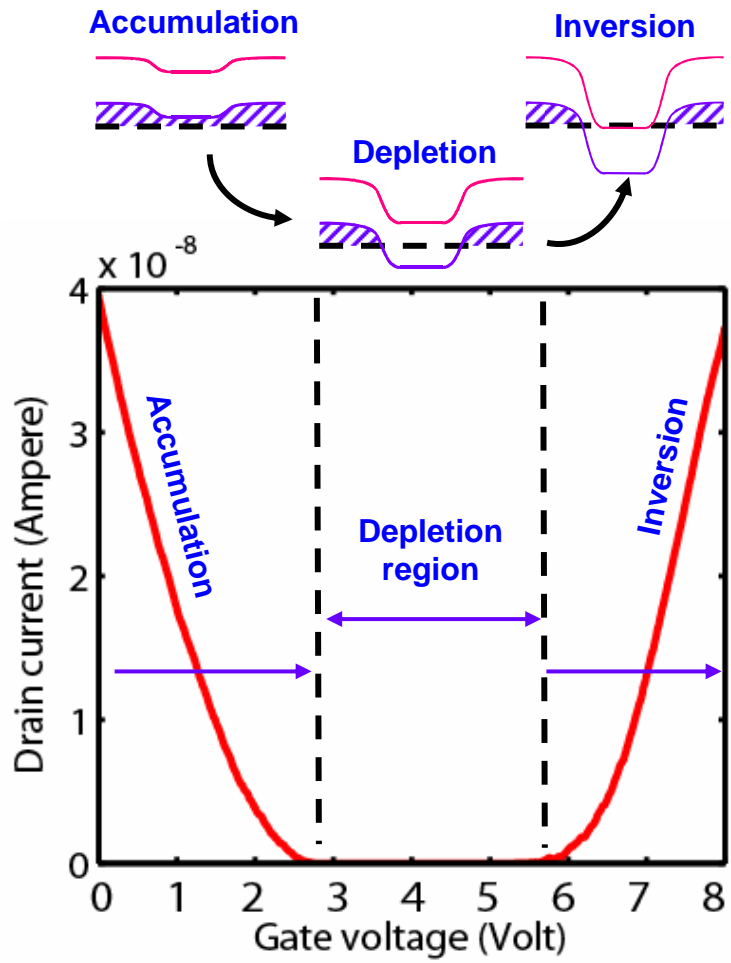

Fig. 3, Jones et al.

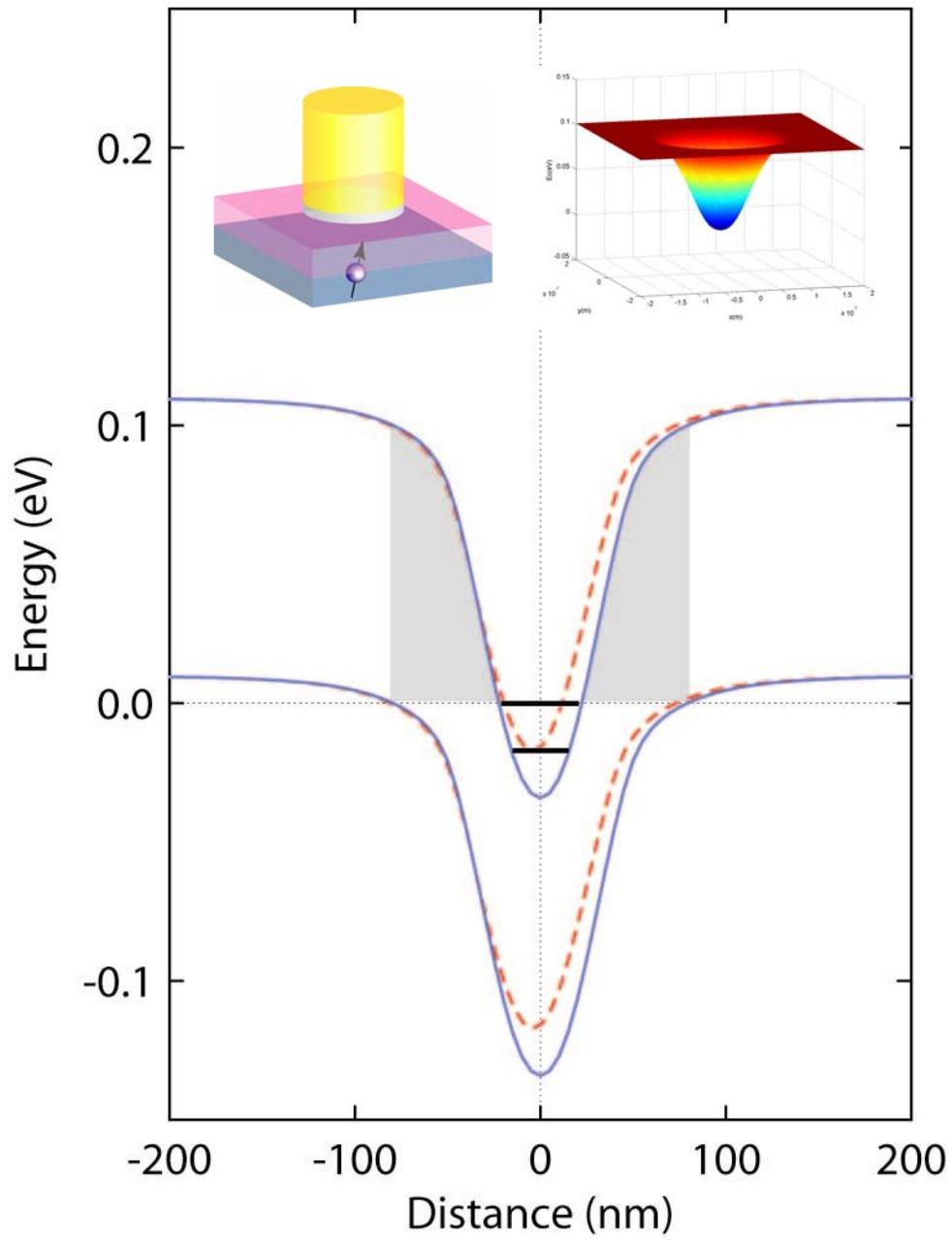

Fig. 4, Jones et al.

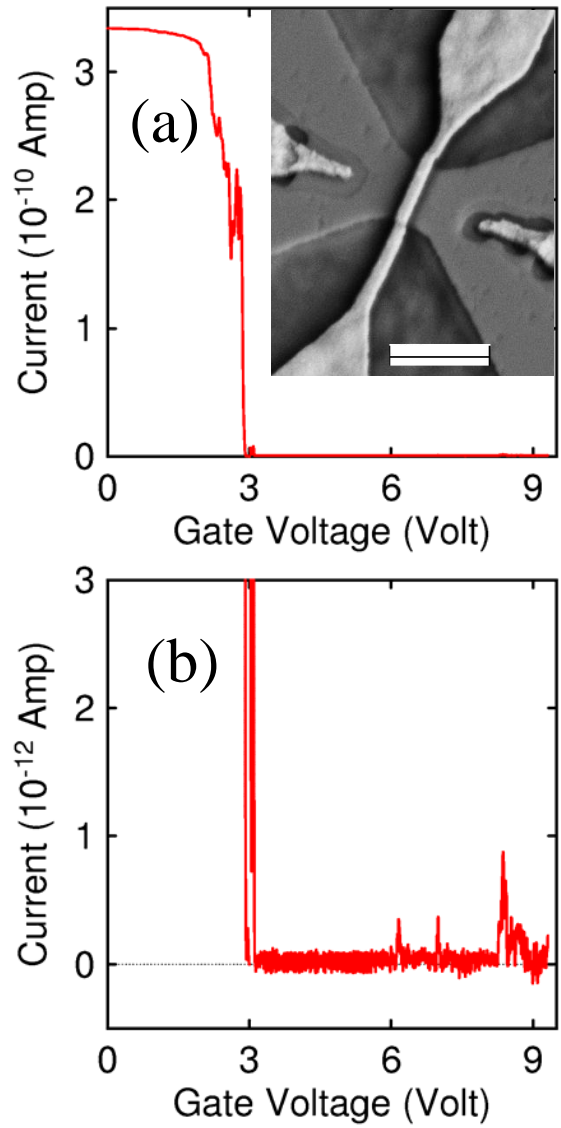

Fig. 5, Jones et al.

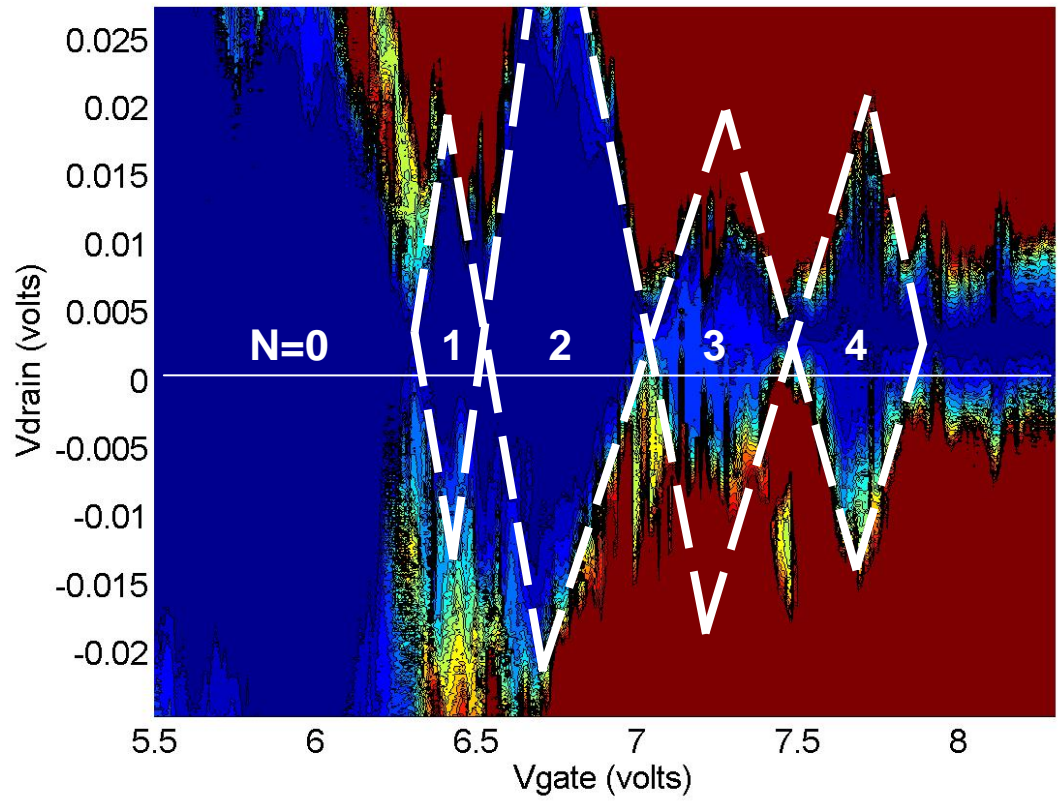